\documentclass[aps,prb,twocolumn,superscriptaddress,showpacs]{revtex4-1}
\usepackage[dvips]{graphicx}
\usepackage{times,color,amsmath}
\usepackage{subfigure}
\usepackage{float}
\usepackage[section]{placeins}

\def  \bxi    {\mbox{\boldmath$\xi $}}
\def  \bbet    {\mbox{\boldmath$\eta $}}
\def  \blam    {\mbox{\boldmath$\lambda $}}

\begin{document}
\title{Two-dimensional electron gas at the LaAlO$_3$/SrTiO$_3$ inteface with a potential barrier}

\author{V. A. Stephanovich}
\affiliation{Institute of Physics, Opole University, ul. Oleska 48, 45-052 Opole, Poland}

\author{V. K. Dugaev}
\affiliation{Department of Physics and Medical Engineering, Rzesz\'ow University of Technology,
al. Powsta\'nc\'ow Warszawy 6, 35-959 Rzesz\'ow, Poland}
\affiliation{Departamento de F\'isica and CeFEMA, Instituto Superior T\'ecnico, Universidade de Lisboa,
av. Rovisco Pais, 1049-001 Lisbon, Portugal}

\author{J. Barna\'s}
\affiliation{Faculty of Physics, Adam Mickiewicz University, ul. Umultowska 85,
61-614 Pozna\'n, Poland}
\affiliation{Institute of Molecular Physics, Polish Academy of Sciences,
60-179 Pozna\'{n}, Poland}

\begin{abstract}

We present a tight binding description of electronic properties of the interface between LaAlO$_3$ (LAO) and SrTiO$_3$ (STO). The description  assumes  LAO and STO perovskites as  sets of atomic layers in the $x$-$y$ plane, which are weakly coupled by an interlayer hopping term along the $z$ axis. The interface is described by an additional potential, $U_0$, which simulates a planar defect. Physically, the interfacial potential can result from either a mechanical stress at the interface or other structural imperfections. We show that depending on the potential strength, charge carriers (electrons or holes) may form an energy band which is localized at the interface and is within the band gaps of the constituting materials (LAO  and STO). Moreover, our description predicts a {\it valve effect} at a certain critical potential strength, $U_{0cr}$, when the interface potential works as a valve suppressing the interfacial conductivity. In other words, the interfacial electrons become dispersionless at $U_0= U_{0cr}$, and thus cannot propagate. This critical value separates the {\it quasielectron} ($U_0<$ $U_{0cr}$) and  {\it quasihole} ($U_0>$ $U_{0cr}$) regimes of the interfacial conductivity.
\end{abstract}

\date{\today}
\maketitle

\section{Introduction}

One of the roads in search for novel materials is based on heterostructures formed at least from two different materials. Such systems can have properties which are significantly  different from those of individual materials. Recently, there has been growing interest in the heterostructures based on oxides, like LaAlO$_3$/SrTiO$_3$ for instance. This is because interfaces between two oxides may have unusual properties.\cite{hwang12,bibes11, heber09}  It has been shown, that even though the constituent oxides are ordinary band insulators with well-known electronic properties,\cite{dagotto11} their interfaces can exhibit a variety  of  phenomena -- from two-dimensional (2D) metallic conductivity,\cite{Ohtomo04,kh12,tiel06,caviglia08,bell09} and superconductivity,\cite{ueno08}  to ferromagnetism~\cite{brinkman07,lee13} and coexistence of magnetic order and superconductivity.\cite{hwang12,lee11,bert11}

Beginning with the paper by Ohtomo and Hwang,\cite{Ohtomo04} the physical properties of the LaAlO$_3$/SrTiO$_3$ interface have been intensively studied in recent years, both experimentally and theoretically. The main peculiarity of such an interface is its perfectly metallic conductivity, with a relatively high electronic mobility, that can be described by the model of 2D electron gas. Moreover, all the above mentioned  unusual properties, like interfacial magnetism and superconductivity occur in this case as well.\cite{bibes11}  Apart from this, anomalous magnetoresistance and Hall effect have been measured in this system, too.\cite{seri09,zhou15,rey12} The emergence of conductive interface is closely related to the metal-insulator transition, which can be realized at this  interface.\cite{kumar15}

Several physical models have been proposed in order to account for  possible origin of the
interface metallicity in the LaAlO$_3$/SrTiO$_3$ system. To our knowledge, one can distinguish three groups of such models. The first group  is related to the so-called polar catastrophe model.\cite{nakagawa06,caviglia08,bell09,tiel06} The second group is based on extrinsic doping by cations such as La$^{3+}$, which is n-type dopant in SrTiO$_3$ (STO).\cite{tok93} The third model, in turn, consists in the formation of bulk-like oxygen vacancies in the STO layers near the interface, which provide free charge carriers.\cite{kal07,sie07,herranz07,syro11,meevasana11,zhuang14}

The {\it polar catastrophe} model is based on the argument that the alternating polarity of atomic layers in LaAlO$_3$ (LAO) along the [001] direction leads to a diverging  electrostatic potential across the structure (hence the words {\it polar catastrophe}), unless the electric charges are reconstructed at the interface. Two possible choices for the connection of LAO and STO impose opposite electrostatic boundary conditions. Namely, LaAlO$_3$ is composed of charged layers of (LaO)$^+$ and (AlO$_2$)$^-$, whereas the corresponding layers in SrTiO$_3$ are electrically neutral. Therefore, terminating the LAO on an atomic plane at the interface breaks the charge neutrality, yielding the above mentioned {\it polar catastrophe} at the interface. To avoid this diverging interface energy, a compensating charge is required. This demonstrates that the {\it polar catastrophe} model is based on the notion of crystal atomic planes which interact with each other along the [001] direction. Below we will use this fact when formulating the tight-binding model.

It should be emphasized that there is still active debate on the possible physical mechanisms of the formation of two-dimensional electron gas,\cite{stemmer14}  the electron energy structure, and the electron states at the LAO/STO interface. Here we present a simplified model, in which the formation of the 2D electron states at the interface is related to a short-range potential barrier (or potential well) near the interface. This potential barrier may be related to a mechanical stress, which is inevitably present at any interface due to mismatch and misfit effects or various kinds of imperfections. In principle, the existence of such barrier at the interface is in agreement with the {\it polar catastrophe} model, which requires the reconstruction of the crystalline structure at the heterointerface in order to avoid the electrostatic potential divergence. In section 2 we describe the tight-binding model of the electronic structure of LAO and also of STO. Section 3 presents description of the electronic states at the interface of STO and LAO. Summary and final conclusions are in section 4.

\section{Electronic spectrum of LAO (STO): tight-binding model}

Having in mind the main objective of this paper, which is to describe electronic properties of the LAO/STO interface, we begin with the model Hamiltonian of a layered crystal like LAO and STO.
To do this we use the approach based on the tight-binding model, with electron hopping within each atomic layer (lying in the
$x$-$y$ plane, see Fig.~1) and  interlayer electron hopping along the axis $z$ (perpendicular to the $x$-$y$ plane). The latter hopping term is assumed to be relatively small and therefore will be considered perturbatively.

\begin{figure}
\includegraphics[width=0.95\columnwidth]{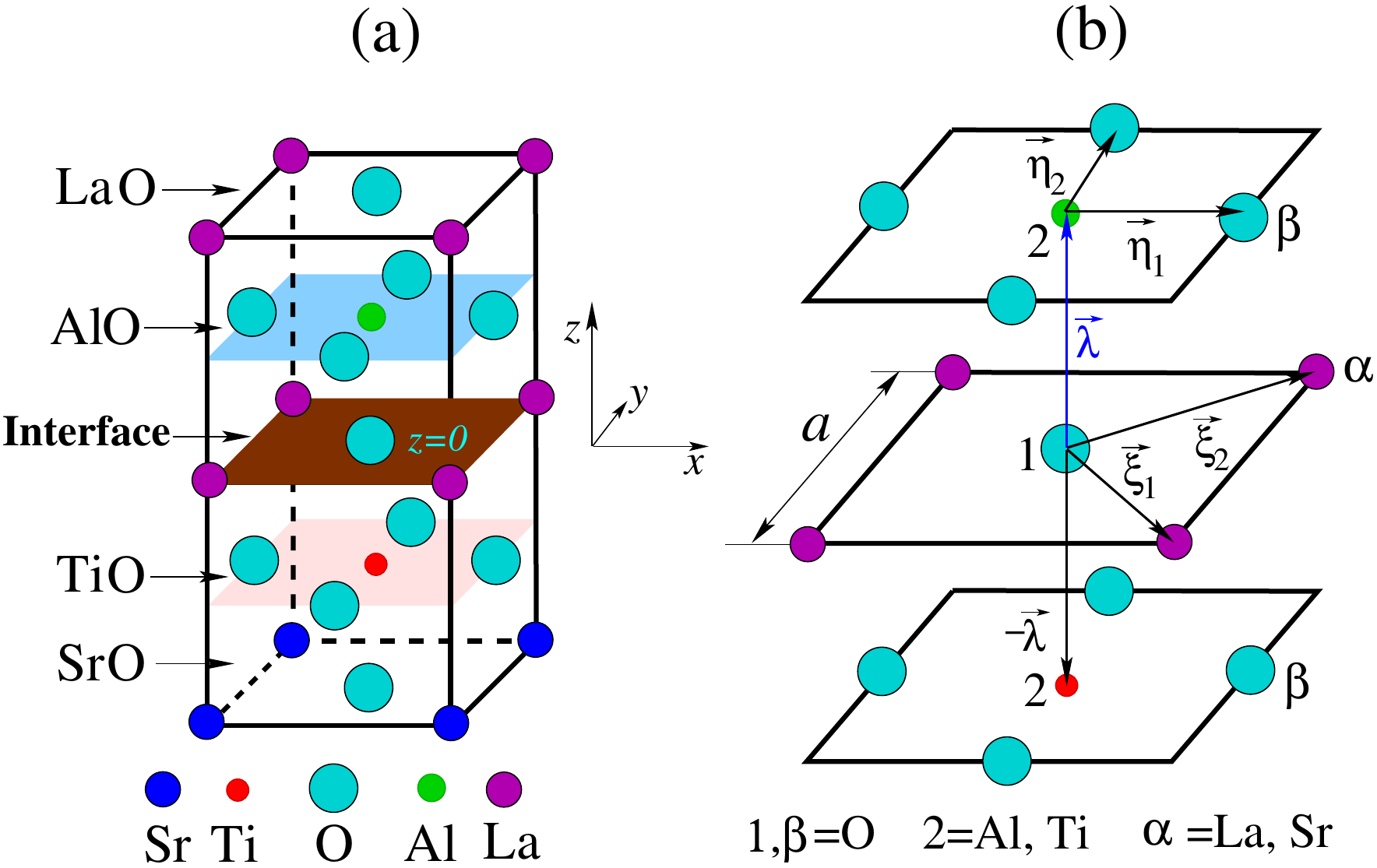}
\caption{(color online) Schematics of LAO-STO interface (a) and geometry of the problem as adopted in the Hamiltonians \eqref{1} and \eqref{3} (b). (b) also shows the local reference frame and vectors describing positions of the corresponding atoms within  the unit cell. The plane $z=0$ corresponds to the interface, while $a$ is the lattice constant.}
\label{f1}
\end{figure}

In the case of LaAlO$_3$, the  intralayer hoping occurs within the La-O and Al-O atomic layers, which are parallel to the $x$-$y$ plane, while the interlayer hopping occurs between Al and O atoms in the neighboring atomic planes, see Fig.~1.
The corresponding Hamiltonian in the basis of atomic orbitals can be written as
\begin{eqnarray}
&&{\cal H}=\sum _{{\bf n}, i}  \varepsilon _i C^\dag _{{\bf n}i} C_{{\bf n}i}
-t_1\sum _{{\bf n},\alpha} C^\dag _{{\bf n}1} C_{{\bf n}\alpha }
\nonumber \\
&&-t_2\sum _{{\bf n},\beta} C^\dag _{{\bf n}2} C_{{\bf n}\beta }
-t_3\sum _{{\bf n}} C^\dag _{{\bf n}1} C_{{\bf n}2} +h.c., \label{1}
\end{eqnarray}
where ${\bf n}=(n_1,n_2,n_3)$ labels crystal unit cells and $i$ refers to the nonequivalent atoms
within the cell. In the following we use the atom's labeling as shown in Fig.1(b). The hopping parameters $t_1$ and $t_2$ are the intra-planar ones, while $t_3$ is the hoping parameter between the layers, i.e. that along the $z$ axis. It is firmly established, that possible metallic conductivity of the LaO/STO interface~\cite{Ohtomo04} occurs in a narrow 2D interface region. Bearing this in mind, our approach to the interface states is to trace how the coupling of  2D atomic layers (due to interlayer hopping, $\sim t_3$) alters their electronic spectrum. To do this we assume  the interlayer hoping term ($t_3$)  as a small perturbation in comparison to the intralayer hopping ($\sim t_{1,2}$).

We use the Fourier transformation
\begin{eqnarray}
\label{2}
C_{{\bf n}i}=\frac1{\sqrt{N}}\sum _{\bf k}e^{i{\bf k\cdot R}_{{\bf n}i}} C_{\bf k} ,
\end{eqnarray}
where ${\bf R}_{{\bf n}i}={\bf R}_{\bf n}+{\bf r}_i$,
with ${\bf R}_{\bf n}=n_1{\bf a}+n_2{\bf b}+n_3{\bf c}$ describing position of the elementary cell, and ${\bf r}_i$ describing position of the $i$-th atom within the cell. Here, ${\bf a, b}$ and $\bf c$  denote the basis vectors of the lattice and $\mathbf k$ is a three-dimensional (3D) wavevector.
The Fourier-transformed Hamiltonian becomes
\begin{eqnarray}
\label{3}
{\cal H}=\sum _{{\bf k}} \Big\{ \sum _{i=1,2,\alpha,\beta}  \varepsilon _i C^\dag _{{\bf k}i} C_{{\bf k}i}\hskip3cm
\nonumber \\
-2t_1\Big[ C^\dag _{{\bf k}1} C_{{\bf k}\alpha }
\Big( \cos ({\bf k}\cdot {\bxi _1})+\cos ({\bf k}\cdot {\bxi _2})\Big) +h.c.\Big]
\nonumber \\
-2t_2\Big[ C^\dag _{{\bf k}2} C_{{\bf k}\beta }
\Big( \cos ({\bf k}\cdot {\bbet _1})+\cos ({\bf k}\cdot {\bbet _2})\Big) +h.c.\Big]
\nonumber \\
-2t_3 \Big[ C^\dag _{{\bf k}1} C_{{\bf k}2} \cos ({\bf k}\cdot {\blam })+h.c.\Big]
\Big\} ,
\end{eqnarray}
where the vectors $\bxi _{1,2}$ and $\bbet _{1,2}$ lie in the $x$-$y$ plane, while
the vector $\blam$ is perpendicular to this plane and oriented along the $z$ axis, see Fig.~1.

In the spirit of the perturbation expansion, we put $t_3=0$ in the zeroth-order of the perturbation scheme. The Hamiltonian \eqref{3} describes then 2D electrons within two inequivalent and decoupled
atomic planes 1 and 2 in the unit cell, with the corresponding electronic spectra
\begin{eqnarray}
E_{1,\pm}({\bf k}_\perp)=\frac{\varepsilon _1+\varepsilon _\alpha }2
\pm \frac12 \Big[ (\varepsilon _1-\varepsilon _\alpha )^2 \hskip2cm \nonumber \\
+16t_1^2\Big( \cos ({\bf k}_\perp\cdot {\bxi _1})+\cos ({\bf k}_\perp\cdot {\bxi _2})\Big) ^2\Big] ^{1/2},
\label{ni4} \\
E_{2,\pm }({\bf k}_\perp)=\frac{\varepsilon _2+\varepsilon _\beta }2
\pm \frac12 \Big[ (\varepsilon _2-\varepsilon _\beta )^2 \hskip2cm \nonumber \\
+16t_2^2\Big( \cos ({\bf k}_\perp\cdot {\bbet _1})+\cos ({\bf k}_\perp\cdot {\bbet _2})\Big) ^2\Big] ^{1/2} \label{ni5},
\end{eqnarray}
where ${\bf k}_\perp$ is the 2D component (in the $x$-$y$ plane) of the wavevector ${\bf k}$.
For definiteness we assume that electrons in the $E_{1,\pm}({\bf k}_\perp)$ subbands have larger energy
than those in the $E_{2,\pm}({\bf k}_\perp)$ ones, and the subbands do not cross each other.
Thus, we can further consider only the closest energy subbands, i.e. $E_{1,-}({\bf k}_\perp)$ and
$E_{2,+}({\bf k}_\perp)$ ones. In the vicinity of the minimum of $E_{1,-}({\bf k}_\perp)$ and maximum of $E_{2,+}({\bf k}_\perp)$, which occur at ${\bf k}_\perp ={\bf 0}$, we can take ${\bf k_\perp}\cdot {\bxi _{1,2}}\ll 1$ and ${\bf k_\perp}\cdot {\bbet _{1,2}}\ll 1$.
Taking now into account the explicit forms of the vectors $\bxi_i$ and $\bbet _i$ in the reference frame of Fig.\ref{f1},
\begin{eqnarray}
&&\bxi _1=\left(\frac a2, -\frac a2\right),\;\; \bxi _2=\left(\frac a2, \frac a2\right), \nonumber \\ \nonumber \\
&&\bbet_1=\left(\frac a2, 0\right), \;\;\bbet_2=\left(0,\frac a2\right), \label{aux2}
\end{eqnarray}
where $a$ is the lattice constant,
we obtain the following long-wavelength expressions for the eigenenergies $E_{1,-}({\bf k}_\perp)$ and $E_{2,+}({\bf k}_\perp)$:
\begin{eqnarray}
\label{uf6}
E_{1,-}({\bf k}_\perp)
\approx \tilde{\varepsilon }_1+\frac{4t_1^2k_\perp^2}{[(\varepsilon _1-\varepsilon _\alpha )^2+64t_1^2]^{1/2}},\\
E_{2,+}({\bf k}_\perp)
\approx \tilde{\varepsilon }_2-\frac{2t_2^2k_\perp^2}{[(\varepsilon _2-\varepsilon _\beta )^2+64t_2^2]^{1/2}}, \label{uf7}
\end{eqnarray}
with
\begin{eqnarray}
\label{8}
\tilde{\varepsilon }_1=\frac{\varepsilon _1+\varepsilon _\alpha }2
-\frac12 \big[ (\varepsilon _1-\varepsilon _\alpha )^2+64t_1^2\big] ^{1/2},
\\
\tilde{\varepsilon }_2=\frac{\varepsilon _2+\varepsilon _\beta }2
+\frac12 \big[ (\varepsilon _2-\varepsilon _\beta )^2+64t_2^2\big] ^{1/2},
\end{eqnarray}
and ${\mathbf k}_\perp$ measured in dimensionless units, ${\mathbf k}_\perp a \Rightarrow {\mathbf k}_\perp$.

The next step is to find the first order  correction to the above spectra of decoupled atomic planes  due to the interlayer hopping term. It is well known \cite{land3} that such a correction is given by the matrix element of the perturbation term calculated with the eigenfunctions of the unperturbed system. Thus, we first calculate the eigenvectors (eigenfunctions)  corresponding to the eigenvalues \eqref{uf6} and \eqref{uf7} at $k_\perp=0$. These eigenfunctions, being  superpositions of the initial electron orbitals $\psi ^{(0)}_1$ and $\psi ^{(0)}_\alpha$ ($\psi ^{(0)}_2$ and $\psi ^{(0)}_\beta$) read
\begin{eqnarray}
\label{10}
\psi _{1,-}=\frac{-4t_1\psi ^{(0)}_1+(\tilde{\varepsilon _1}-\varepsilon _1)\, \psi ^{(0)}_\alpha }
{\Bigl[16t_1^2+(\tilde{\varepsilon _1}-\varepsilon _1)^2\Bigr] ^{1/2}},
\\
\psi _{2,+}=\frac{-4t_2\psi ^{(0)}_2+(\tilde{\varepsilon _2}-\varepsilon _2)\, \psi ^{(0)}_\beta }
{\Bigl[ 16t_2^2+(\tilde{\varepsilon _2}-\varepsilon _2)^2\Bigr] ^{1/2}}. \label{10a}
\end{eqnarray}
In view of Hamiltonian \eqref{1}, hopping between these states is related to overlapping of the $\psi ^{(0)}_1$ and $\psi ^{(0)}_2$ orbitals. The corresponding correction to the energy is then
\begin{equation}
\delta E=
-2 \tilde t_3 \cos(k_z\lambda),\label{xau1}
\end{equation}
where
\begin{equation}
\tilde t_3=\frac{16t_1t_2t_3}{\Big\{ \Big[ 16t_1^2+(\tilde{\varepsilon _1}-\varepsilon _1)^2\Big]
\Big[16t_2^2+(\tilde{\varepsilon _2}-\varepsilon _2)^2\Big] \Big\} ^{1/2}}.
\end{equation}
When deriving Eq.\eqref{xau1} we took into account the fact that only the term $\sim \psi ^{(0)}_1 \psi ^{(0)}_2$ is nonzero.

Thus, to describe electrons in the considered electronic bands, we can restrict ourselves to an effective 3D model which includes
electron hopping between the 2D electron bands. In the matrix form this Hamiltonian can be written as
\begin{eqnarray}
\label{12}
 {\mathcal H}_{\rm eff} =\sum _{\bf k}\Psi ^\dag _{\bf k}
\left( \begin{array}{cc}
E_{1,-}({\bf k}_\perp) & -2\tilde{t}_3\cos (k_z\lambda ) \\
-2\tilde{t}_3\cos (k_z\lambda) & E_{2,+}({\bf k}_\perp)
\end{array} \right) \Psi _{\bf k}. \hskip0.3cm
\end{eqnarray}
The smallest energy distance of the electron bands in the $z$ direction occurs at $k_z\lambda =\pi /2$. Thus, the point ${\bf k}^0=(0,0,\pi /2\lambda )$ determines
the energy gap $\Delta =\tilde{\varepsilon _1}-\tilde{\varepsilon _2}$.
 Hamiltonian describing  electronic states in the vicinity of this point can be written as
\begin{eqnarray}
\label{14}\
\mathcal{H}=\left( \begin{array}{cc}
\Delta +k_\perp^2/2m_c & vk_z \\
vk_z & -\Delta -k_\bot^2/2m_v
\end{array} \right),
\end{eqnarray}
where we shifted the zero energy point to the middle of the gap, and $k_z$ is measured from the point $k_z^0=\pi /2\lambda$ and is  in dimensionless units, similarly as ${\bf k}_\perp$. The parameters $m_{c,v}$
have been  introduced to simplify the notations [these parameters can be determined by comparing Eq.\eqref{14} with Eqs (7) and (8)], and are related to the corresponding effective masses. In turn,  the parameter $v$ is defined as $v=2\tilde{t}_3\lambda/a $.

\section{Junction with a potential barrier at the interface}

\subsection{Equation for the interface electronic states}

Now we use the model Hamiltonian \eqref{14} to describe a junction of two similar materials, which differ
in the parameters $\Delta $, $m_c$, $m_v$, and $v$. In the following these materials are distinguished with
the index 1 and 2. Apart from this, we assume a potential barrier $U_0$ at the interface $z=0$, and a band offset
equal to $U_1$. Thus, the potential  profile across the structure is
\begin{eqnarray}
\label{15}
U(z)=U_0\, \delta (z)+U_1\theta (z)
\end{eqnarray}
and $U(z)=0$ for $z<0$, where $\theta(z)$ is the Heaviside function.
Taking into account Eq.\eqref{14} one can write the Hamiltonian for  such a junction in the form
\begin{eqnarray}
\label{16}\
\mathcal{H}=\left( \begin{array}{cc}
s_c+U & -iv\nabla_z \\
-iv\nabla _z & -s_v+U
\end{array} \right) ,
\end{eqnarray}
where $\nabla_z=\partial/\partial z$ (note $z$ and $\nabla_z$ are here dimensionless, $z/a\Rightarrow z$, and we put $\hbar =1$),  and
\begin{subequations}
\begin{align}
\label{17}
s_c(z)=s_{c1}[1-\theta (z)]+s_{c2}\theta (z), \\
s_v(z)=s_{v1} [1-\theta (z)]+s_{v2}\theta (z),\\
v(z)=v_1[1-\theta (z)]+v_2\theta (z),
\end{align}
\end{subequations}
with
\begin{equation}
s_{c1,2}=\Delta_{1,2}+\frac{k_\perp^2}{2m_{c1,2}},\
s_{v1,2}=\Delta_{1,2}+\frac{k_\perp^2}{2m_{v1,2}}. \label{17a}
\end{equation}

The Schr\"odinger equation with the Hamiltonian \eqref{16} for the spinor components $\varphi ,\chi $
of the wavefunction reads
\begin{subequations}
\begin{align}
(s_c+U-\varepsilon )\, \varphi -iv\nabla _z\chi =0,\\
-iv\nabla _z\varphi +(-s_v+U-\varepsilon )\, \chi =0.
\label{21}
\end{align}
\end{subequations}
From Eq.\eqref{21} follows that
\begin{eqnarray}
\label{22}
\chi =\frac{iv}{-s_v+U-\varepsilon }\, \nabla _z\varphi .
\end{eqnarray}
Substituting Eq.(22) into Eq.(21a) we find the following equation for $\varphi (z)$:
\begin{eqnarray}
\label{23}
(s_c+U-\varepsilon )\, \varphi +v\nabla _z\, \frac{v}{-s_v+U-\varepsilon }\, \nabla _z\varphi =0.
\end{eqnarray}

We look now for solutions localized at the interface.
For $z<0$, there is a solution $\varphi (z)=Ae^{\kappa _1z}$. Substituting this solution into Eq.(23)
we find
\begin{eqnarray}
\label{24}
\kappa _1=\frac1{v_1} [(s_{c1}-\varepsilon )(s_{v1}+\varepsilon )]^{1/2}.
\end{eqnarray}
Similarly, for $z>0$ there is a solution $\varphi (z)=Ae^{-\kappa _2z}$,
which fulfills the continuity condition at $z=0$, and
\begin{eqnarray}
\label{25}
\kappa _2=\frac1{v_2} [(s_{c2}+U_1-\varepsilon )(s_{v2}-U_1+\varepsilon )]^{1/2}.
\end{eqnarray}
Note that if we take $k_\bot =0$ and assume $U_1>0$,\cite{note} then $\kappa _1$ and $\kappa _2$ from Eqs.~\eqref{24} and \eqref{25} are real provided the following inequalities $U_1-\Delta _2<\varepsilon <U_1+\Delta _2$ and $-\Delta _1<\varepsilon <\Delta _1$ are fulfilled simultaneously. Their detailed analysis will be presented below.

Dividing Eq.\eqref{23} by $v(z)$ and integrating over a small vicinity near the interface ($z=0$), we arrive at the
following equation for the energy of electron states localized at the interface:
\begin{equation}
Q+\frac{v_2\kappa _2}{s_{v2}-U_1+\varepsilon} +\frac{v_1\kappa _1}{s_{v1}+\varepsilon}=0,
\label{26}
\end{equation}
where $Q$ is defined as
\begin{equation}\label{e27}
Q=\frac{U_0}{2}\left(\frac{1}{v_1}+\frac{1}{v_2}\right).
\end{equation}
Taking into account the definitions of $\kappa_{1,2}$ (see Eqs \eqref{24} and \eqref{25}), this equation can be reduced to the form
\begin{eqnarray}
\label{27}
Q+\left( \frac{s_{c2}+U_1-\varepsilon }{s_{v2}-U_1+\varepsilon }\right) ^{1/2}
+\left( \frac{s_{c1}-\varepsilon }{s_{v1}+\varepsilon }\right) ^{1/2}=0.
\end{eqnarray}
Note that this equation can have real solutions only when $U_0<0$, or equivalently $Q<0$, see Eq.\eqref{e27}.  The  equation \eqref{27} is the main result of this paper as it defines the spectrum of charge carriers (electrons or holes), localized at the interface [in the $z$ direction, see the expressions \eqref{24} and \eqref{25}], and delocalized in the interface plane ($x$-$y$ plane). Furthermore, the conditions for localization in the $z$ direction ($\kappa_{1,2}$ are real) ensure that the entire spectrum \eqref{27} is inside the band gaps of the constituting materials.
Before solving Eq.\eqref{27}, it is worth to present explicitly the equation describing energy  $\varepsilon _0$ of the localized states with $k_\perp=0$.
In this case $s_{c1}=$ $s_{v1}=$ $\Delta_1$, $s_{c2}=$ $s_{v2}=$ $\Delta_2$, and from Eq.\eqref{27} one finds
\begin{eqnarray}
\label{28}
Q+ \left( \frac{\Delta _2+U_1-\varepsilon _0}{\Delta _2-U_1+\varepsilon _0}\right) ^{1/2}
+\left( \frac{\Delta _1-\varepsilon _0}{\Delta _1+\varepsilon _0}\right) ^{1/2}=0.\hskip0.6cm
\end{eqnarray}

\subsection{Spectrum of electronic states localized at the interface: symmetrical case}

Analytical solution of Eq.\eqref{27} is possible for a symmetrical case, i.e. when the band offset $U_1=0$, both materials have equal gaps, $\Delta_1= \Delta_2= \Delta$, and equal effective masses in both valence and conduction bands, $s_{c1}=s_{c2}=s_{c}$,  $s_{v1}=s_{v2}=s_{v}$, and also $v_1=v_2=v$. Then,  from Eq.\eqref{27} we obtain
\begin{equation}\label{sol1}
Q+2\sqrt{\frac{s_c-\varepsilon}{s_v+\varepsilon}}=0,
\end{equation}
where now $Q=U_0/v$. Bearing in mind that $U_0<0$ and $Q<0$, the above equation has a solution
\begin{equation}\label{sol2}
\varepsilon({\bf k}_\perp)=\frac{s_c-\gamma U_0^2s_v}{1+\gamma U_0^2},
\end{equation}
where  $\gamma=1/4v^2$.
Taking $k_\perp=0$ in Eq.\eqref{sol2}, one finds the solution of Eq.\eqref{28},
\begin{equation}\label{sol3}
\varepsilon_0=\Delta \frac{1-\gamma U_0^2}{1+\gamma U_0^2}.
\end{equation}
From Eq.\eqref{sol2} follows that the localized electronic band $\varepsilon({\bf k}_\perp)$ exists for any $U_0<0$; also at $U_0\to 0$, when  $\varepsilon({\bf k}_\perp) \to s_c$, and at $U_0\to -\infty$, when $\varepsilon({\bf k}_\perp) \to -s_v$. This behavior reflects the physics of the system. At $k_\perp=0$, see Eq.\eqref{sol3}, these asymptotics read:
$\varepsilon _0\to \Delta $ for $U_0\to 0$, and   $\varepsilon _0\to -\Delta $ for $U_0\to -\infty$.

Let us introduce dimensionless variables:
$ \varepsilon / \Delta= E$ and $q_{c,v}=\mu_{c,v} / \Delta$, with $\mu_{c,v}=1/2m_{c,v}$.
Taking into account the fact that now $Q=U_0/v$, the  dimensionless version of the solution \eqref{sol2} can be written as
\begin{equation}
E({\mathbf k}_\perp )=\frac{1-Q^2/4+k_\perp^2(q_c-q_vQ^2/4)}{1+Q^2/4},  \label{res5}
\end{equation}
One can easily check that the spectrum $E(k_\perp)$ given by Eq.\eqref{res5} is bounded between two curves corresponding to $Q=0$ and $Q \to -\infty$. Indeed, one can  see from Eq.\eqref{res5} that  $E(Q=0)=1+q_ck_\perp^2$, which corresponds to {\it quasielectron} type of interfacial conductivity, and  $E(Q \to -\infty)=-(1+q_vk_\perp^2)$, which can be ascribed to  {\it quasihole} type of conductivity. In the dimensional variables these limiting curves can be written as $\varepsilon=\pm \Delta (1+q_{c,v}k_\perp^2)$. This reflects simply the fact that the interfacial spectrum is located inside the gap of the constituting materials (which so far is the same for both materials). In the subsection C  we will see that this also holds in a more general (asymmetric) case,  Eq.\eqref{res3}.

As $|Q|$ grows, the energy $E(k_\perp =0)$ becomes gradually negative, but more important is the fact that the coefficient in front of  $k_\perp^2$ becomes zero at $Q_{cr}=-2\sqrt{q_c/q_v} = -2\sqrt{m_v/m_c}$, and then it becomes negative with a further increase in $|Q|$. This is equivalent to inverting of the parabola and thus changing the type of conductivity from quasielectron to quasihole. Thus, the energy spectrum becomes dispersionless for $Q=Q_{cr}$, so  the corresponding quasiparticles cannot propagate, making the interface  nonconductive. This also means that the interface potential $U_0$ works as a {\it valve} closing (at the critical value $U_{0cr}$) the electron current in the LAO/STO interface. As the critical value $U_{0cr}$ separates the {\it quasielectron} and {\it quasihole} regimes of interfacial conductivity, we can suggest that variation of the potential $U_0$ can not only control the type of interfacial conductivity (electron or hole), but also can make the interface nonconducting  at $U_0=$ $U_{0cr}$.
In dimensional variables, the corresponding expression for $U_{0cr}$ takes the form
\begin{equation} \label{res6}
U_{0cr}=-2v\sqrt{\frac{m_v}{m_c}}.
\end{equation}

\begin{figure}
\includegraphics[width=0.95\columnwidth]{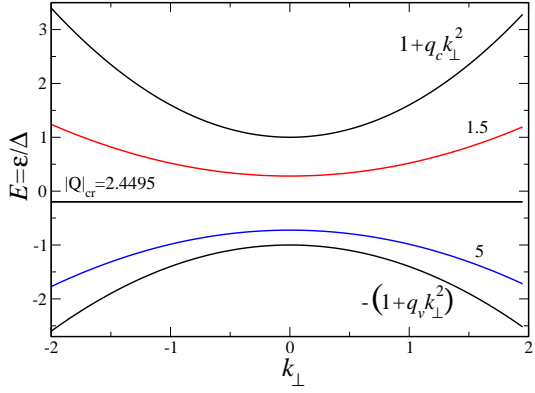}
\caption{The plot of spectrum described by Eq.\eqref{res5} for $q_c=0.6$ and $q_v=0.4$. The limiting curves, $\pm (1+q_{c,v}k_\perp^2)$, and the curve corresponding to the critical value $Q_{cr}=-|Q_{cr}|$ are also shown. The numbers at the other curves show the corresponding values of $|Q|$.}
\label{f3}
\end{figure}

The spectrum \eqref{res5} is presented in Fig.\ref{f3} for  $q_c=0.6$ and $q_v=0.4$. This figure  visualizes the above discussed qualitative behavior in the {\it quasielectron}, {\it quasihole}, and nonconductive (at $Q=Q_{cr}$) regimes. Furthermore, the spectrum is bounded by the curves $\pm(1+q_{c,v}k_\perp^2)$.
Below we shall see that the situation is similar in  a more general case, i.e. in the {\it asymmetric} one. The only difference is the existence of a threshold value $Q_{tr}$ of the parameter $Q$,  such that Eq.\eqref{27}  does not have solutions localized at the interface for
$|Q|<|Q_{tr}|$.

\subsection{Solution of the equation for interfacial spectrum in an asymmetrical case}

To investigate equation \eqref{27} for the electronic spectrum in a general (asymmetrical) case, it is convenient to introduce the following dimensionless parameters
\begin{equation} \label{res2}
\frac{\Delta_2}{\Delta_1}=\delta,\ \frac{U_1}{\Delta_1}=u_1,\ \frac{\varepsilon}{\Delta_1}=E,\ q_l=\frac{\mu_l}{\Delta_1},\ \mu_l=\frac{1}{2m_l},
\end{equation}
where we denote $l=c1, c2, v1,  v2$.

Using the dimensionless variables defined above, see Eq.~\eqref{res2}, one can rewrite the equation \eqref{27} in the form
\begin{equation} \label{res3}
Q+\sqrt{\frac{\delta+u_1-E+q_{c2}k_\perp^2}{\delta-u_1+E+q_{v2}k_\perp^2}}+\sqrt{\frac{1-E+q_{c1}k_\perp^2}{1+E+q_{v1}k_\perp^2}}=0.
\end{equation}
For completeness, we also list here the dimensionless version of Eq.~\eqref{28}, which also can be obtained from Eq.~\eqref{res3} simply by putting $k_\perp =0$,
\begin{equation} \label{res4}
Q+\sqrt{\frac{\delta+u_1-E_0}{\delta-u_1+E_0}}+\sqrt{\frac{1-E_0}{1+E_0}}=0,
\end{equation}
with  $E_0=\varepsilon_0/\Delta_1$.
To have real solutions, both expressions under the square roots in Eq.\eqref{res4} should be non-negative. This condition is equivalent to the conditions for having  real  parameters
$\kappa_{1,2}$ in Eqs. \eqref{24}   and \eqref{25}, see the comment below  Eq. \eqref{25}, and is satisfied if the energy $E_0$ obeys the inequalities
\begin{equation}\label{ra}
{\rm max}\, (-1,u_1-\delta)\le E_0\le {\rm min}\, (1,u_1+\delta).
\end{equation}
The criterion (38) also implies that $u_1-\delta <1$; otherwise there is no real solution for $E_0$.

\begin{figure}
\includegraphics[width=0.95\columnwidth]{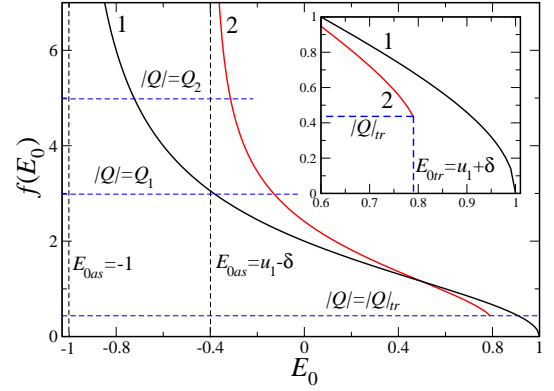}
\caption{Plot of the function given by Eq.\eqref{res7} for $\delta=1$, $u_1=0$ [curve 1, see also the expression \eqref{sol3}] and for $\delta=0.6$, $u_1=0.2$ (curve 2). The roots of the equation $|Q|=f(E_0)$ are defined by the intersection points of the curves 1 and 2 with the horizontal dashed lines corresponding to different values of $|Q|$. For the curve 2 the roots exist for $|Q|\ge|Q_{tr}|$, while for the curve 1 they exist for all $|Q|$'s. The vertical asymptotes define the roots at $|Q| \to \infty$ for the curves 1 ($E_{0as}=-1$) and for the curve 2 ($E_{0as}=u_1-\delta$). These asymptotes correspond to zero of the denominator in the first term of Eq.\eqref{res7}. The inset shows the vicinity of $|Q_{tr}|$. Here, $E_{0tr}$ corresponds to the zero of the numerator in the first term of Eq.\eqref{res7}.}
\label{f4}
\end{figure}

The main qualitative difference between the spectra in the asymmetric and  symmetric (Eq.\eqref{res5}) cases is the existence of a certain threshold value $Q_{tr}$ in the former case. One may expect (and our numerical calculations confirm this) that if the solution of equation \eqref{res3} exists for $k_\perp =0$, then it exists for all values of $k_\perp $. This means that equation \eqref{res4} for $k_\perp =0$ is sufficient to determine $Q_{tr}$.
To do this let us  define the following function:
\begin{equation} \label{res7}
f(E_0)=\sqrt{\frac{\delta+u_1-E_0}{\delta-u_1+E_0}}+\sqrt{\frac{1-E_0}{1+E_0}},
\end{equation}
so that the equation \eqref{res4} can be written as $|Q|=f(E_0)$ (we remind that $Q$ is negative, $Q=-|Q|$). This equation will be  solved graphically as shown in Fig.\ref{f4}. The roots of the equation $|Q|=f(E_0)$ are the abscissas of the intersection points of the curves $f(E_0)$ and the horizontal straight dashed lines corresponding to different values of $Q$. We see that the curves $f(E_0)$ are in a limited energy region -- from   $E_{0cr}$, determined by zero of the numerator of the first square root in Eq.\eqref{res4}, to $E_{0as}$ determined by zero of the denominator of the first square root in Eq.\eqref{res4}. This implies that the solution of equation $|Q|=f(E_0)$ [it is actually the reciprocal function of $f(E_0)$, see \eqref{res7}] is also limited by the above values of $E_0$. Thus,  the resulting spectrum $E(k_\perp)$ of Eq.\eqref{res3} is bounded by two limiting curves. This behavior is similar to that in the {\it symmetric case}, Eq.\eqref{res5}.

Now, we can determine $Q_{tr}$ from Eq.\eqref{res4} analytically. To do this, we recall first that the sum of two square roots is always positive, $\sqrt{a}+\sqrt{b}>0$; it can be zero if and only if $a=0$ and $b=0$.
From Fig. \ref{f4} follows that $|Q_{tr}|$ obeys the equality $|Q(E_{0tr})| = $ $f(E_{0tr})$. On the other hand, $E_{0tr}=\delta+u_1$ corresponds to zero of the numerator in the first term of \eqref{res7}.
Thus, as the first term of \eqref{res7} is zero at $E_0=E_{0tr}$ , the determination of $|Q_{tr}|$ is reduced to the substitution of $E_{0tr}=\delta+u_1$ to the second term. This yields
 \begin{equation} \label{res10}
|Q_{tr}|\equiv f(E_{0tr}) =\sqrt{\frac{1-u_1-\delta}{1+u_1+\delta}}.
 \end{equation}
 The radicand of \eqref{res10} is positive if
  \begin{equation} \label{res11}
-1 \leq u_1+\delta \leq 1.
 \end{equation}

\begin{figure}
\includegraphics[width=0.95\columnwidth]{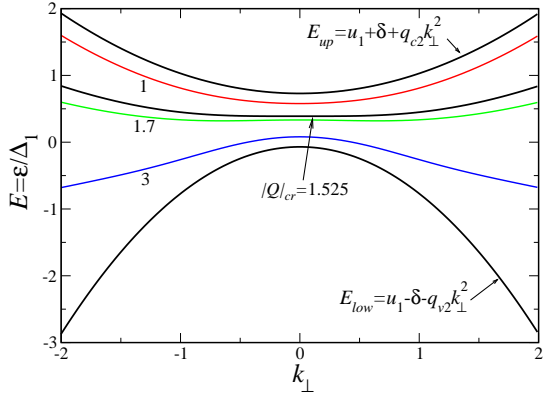}
\caption{The plot of numerical solution of Eq. \eqref{res3} for the spectrum. The parameter values:
$\delta=0.4$, $u_1=0.33$, $q_{c1}=0.6$, $q_{c2}=0.3$, $q_{v1}=0.2$, $q_{v2}=0.7$. These values yield
$|Q_{tr}|\approx 0.3951$ (corresponding to limiting curve $E_{up}(k)$) and $|Q_{cr}|\approx 1.525$.
Limiting curves $E_{up}(k)$, $E_{low}(k)$ \eqref{po1} and that for critical value $|Q_{cr}|$ are shown. Figures near curves correspond to $|Q|$ values.}
\label{f5}
\end{figure}
The expressions \eqref{res10} and \eqref{res11} determine the desired threshold for the existence of the solution of Eq. \eqref{res3}. In dimensional variables they yield
\begin{eqnarray}
&&U_{0tr}=-\frac{2v_1v_2}{v_1+v_2}\sqrt{\frac{\Delta_1-U_1-\Delta_2}{\Delta_1-U_1+\Delta_2}}, \label{res12} \\ \nonumber \\
&& -\Delta_1 \leq U_1 +\Delta_2 \leq \Delta_1.  \label{res13}
\end{eqnarray}
One can see that $U_{0tr} \neq 0$ either at nonzero band offset $U_1 \neq 0$ or if the energy gaps of the constituting materials are not equal, $\Delta_1 \neq \Delta_2$.

The requirement of positive radicands in Eq.\eqref{res3} yields the upper $E_{up}(k_\perp)$ and lower $E_{low}(k_\perp)$ bounds for its solutions $E(k_\perp)$:
\begin{subequations}
\begin{align}
&E_{low}(k_\perp) \leq E \leq E_{up}(k_\perp),\label{po1} \\
&E_{up}(k_\perp)=\delta +u_1+q_{c2}k_\perp^2,  \\
&E_{low}(k_\perp)=u_1-\delta-q_{v2}k_\perp^2.
\end{align}
\end{subequations}
It is seen that $E_{up}(k_\perp)$ corresponds to the {\it quasielectron} type of conductivity, while $E_{low}(k_\perp )$ to the {\it quasihole} one.

The determination of $|Q_{cr}|$ at which the spectrum becomes dispersionless is a little more involved than it was  in the symmetric case \eqref{res5}. The natural condition here is to find $Q$ at which the coefficient in front of $k_\perp^2$ in the solution $E(k_\perp)$ of \eqref{res3} vanishes. This coefficient is obtained from the Taylor expansion of the square roots in Eq.\eqref{res3} up to $k_\perp^2$. The resulting algorithm is as follows. First, we determine the $E_0=E_{cr}$ from the equation
\begin{widetext}
\begin{equation}\label{po2}
\frac{q_{c1}(1+E_{cr})-q_{v1}(1-E_{cr})}{2(1-E_{cr}^2)}\, \sqrt{\frac{1-E_{cr}}{1+E_{cr}}}+ \frac{q_{c2}(E_{cr}-u_1+\delta)-q_{v2}(\delta-E_{cr}+u_1)}{2\left[\delta^2-(E_{cr}-u_1)^2\right]}\,\sqrt{\frac{u_1+\delta-E_{cr}}{E_{cr}-u_1+\delta}}=0.
\end{equation}
\end{widetext}
which is indeed the coefficient at $k_\perp^2$ in the above Taylor expansion. Then, $|Q_{cr}|$ is determined from the condition $|Q_{cr}|=f(E_{cr})$, where $f(E_{cr})$ is given by Eq. \eqref{res7}. Note that in the {\it symmetric} case, Eq.\eqref{sol2}, the expression \eqref{po2} yields $E_{cr}=(1-z)/(1+z)$, $z=q_c/q_v$, which,
when being substituted to $f(E)$ \eqref{res7} gives exactly the result \eqref{res6}.

In Fig. \ref{f5} we show the solutions of Eq.\eqref{res3} for some representative parameters. This figure is qualitatively similar to Fig. \ref{f3} with a few exceptions. First,  the limiting curve $E_{up}(k_\perp)$ corresponds now to a threshold value $|Q_{tr}|\approx 0.3951$ of the parameter $Q$ rather then to $Q=0$. Other difference is that at $|Q|=|Q_{cr}|\approx 1.525$, the spectrum $E(k_\perp)$ is dispersionless only in the central part, which makes sense in our problem as we are using long-wavelength approximation. Hence, the long wavelength behavior of $E(k_\perp)$ is qualitatively similar to that in the {\it symmetric} case. except that now we have the threshold values of the interface potential $Q$.

\section{Conclusions}

In summary, the tight-binding model we applied to a junction with a potential barrier at the interface is a natural continuation of a microscopic description of layered oxide structures, stemming from the paper by Ohtomo and Hwang,\cite{Ohtomo04}  see Fig.1 of this paper. The natural way to model the mechanical stresses and crystalline structure imperfections, which is inevitably present at the interface, is to introduce the interfacial potential \eqref{15}, which can modify the electronic structure of the LAO/STO heterojunction. We have shown that, depending on the potential strength $U_0$, the interfacial conductivity can change its character from $n-$ (quasielectron)  to $p-$type  (quasihole) with some threshold value $U_{0tr}$ \eqref{res12}, at which the charge carrier becomes dispersionless and thus cannot propagate. This means that at some interfacial potential strength, $U_0=U_{0tr}$, this potential (related to mechanical stress and/or imperfections) works as a {\it valve}, which suppresses  the interfacial conductivity and also separates the regions of $n-$type ($U_0<U_{0cr}$) and $p-$type ($U_0>U_{0cr}$) conductivity. In other words, the variation of the interfacial potential can  modulate the conductivity, which may be used in the designing of functional interfaces for oxide electronic devices.
Of course, for real interfacial conductivity to occur, our interface band  should be filled by electrons or holes. The criterion is $E_0<E_F$ ($E_F$ is Fermi level) for $n$-type and  $E_0>E_F$ for $p$-type, where $E_0$ is parabola vertex defined by the solution of Eq. \eqref{res4}.

An important feature of the LAO/STO interface is the strong sensitivity of its transport properties to electric field. This field can be either external or induced, for example, by ferroelectric polarization of  additional layers of Pb(ZrTi)O$_3$  (PZT).~\cite{bhalla10,tra13,stengel11,Shkl15}.
Finally, we have demonstrated that the interfacial potential related to the mechanical stresses and/or defects can control the conductivity of the LAO/ STO interface, changing its type from quasielectron to quasihole. In order to gain better insight into the fundamental mechanism behind this intriguing behavior, it is crucial to perform further theoretical and experimental studies of electronic structure at the LAO/STO heterointerface.

\begin{acknowledgments}
This work was supported by the National Science Center in Poland as a research project
No.~DEC-2012/06/M/ST3/00042.
\end{acknowledgments}


\begin{thebibliography}{99}
\bibitem{note}
The condition $U_1>0$ can be used without loss of generality as long as we do not
impose any additional condition for the ratio $\Delta _2/\Delta _1$.
\end{thebibliography}
\end{document}